# Prosthetics of the Indian State

The e-Shram Portal for Unorganized Workers in India

Rozin Hasin

12567582

Media Theories Seminar – Final Paper

MA New Media & Digital Culture

Supervision: Dr. Niels v. Doorn

Department of Media Studies, University of Amsterdam

20th December, 2024

3159 words

# Table of Contents

# Prosthetics of the Indian state: The e-Shram portal for unorganized workers

The developments of the platform economy, or "gig work" is one the most pertinent topics complicating current debates about work and labor regulations (Faraoun, 2024). In the Global South, particularly South-East Asia, researchers have shown that contrary to the public messaging of platforms of the new forms of service and gig work such as ride-hailing, food delivery, or beauty/household services as a lucrative source of supplementary income; a vast section of the labor force working in today's modern platform economy do so because it one of the few forms of continuous (yet increasingly precarious) employment opportunities available to them (Katta et al., 2024; Nair, 2023; Ray, 2024).

The COVID 19 lock-downs accelerated the platformization of our daily lives, and deep-rooted the desires of the middle class for increased personalization of their daily services, through the ubiquitous proliferation of super-app ecosystems and access to better Internet services. However, the same rise in platformization has worsened the labor conditions of the workers engaged in service-work platforms – ride-sharing, food and essential goods delivery, and household/beauty services. While in urban centres, Zomato and Uber workers were hailed as society's heroes, investigative reports were being published on countless deaths of migrant workers en-route to their villages after they were just fired from their sub-contracted jobs. The corporate media assisted the state in covering up the deaths of these laborers, some of which happened at the hands of the state itself, trying to institute lockdown restrictions with border policies in the scorching heat during summer.

The central government released a press report titled "Migration in India 2020-21" (PIB Delhi, 2022) which provides some interesting and conflicting insights into how people moved to cope with the state of the society after one of the heaviest pandemic lock-downs in March, 2020. According to this report, around 30% of both rural and urban males moved between states, while it was lower for rural and urban females (4 and 14.9% respectively; 2022). One of the leading investigative digital publications in the country, TheWire.in, incidentally misreported this statistic in the second paragraph of a piece published by their editorial staff one week after the governments release, where they picked the first number from the column on rural males migrating from urban areas in other states, reporting that "51.6% of migrants moved from urban areas" (The Wire Staff, 2022). While the visualization's later in the piece reflect the statistics accurately from the report, this exaggerated number in the introduction mars the critical promise of such journalistic endeavors. Effects of the pandemic have been studied more judiciously in academic settings, particularly in sociological research, and research speculating new frameworks of labor and health policy (Carswell et al., 2022; Sinha & Pandit, 2023; Srivastava, 2020a; Yadav & Priya, 2021). The socio-cultural, and the closely linked political, structures make India a unique case for internal migration and circular labor forces, due to the diversity of its regional languages, local labor regimes augmented by state-corporate relations, and barriers to social security and reproductive labor rights essential for migrant lives (Srivastava 2020b, p. 12; Carswell et al., 2022, p. 4605). Arguing on how this internal migration works, Srivastava argues that the contemporary exploitation of internal migrants is rooted in uneven development across the country, which leads to circular migration from "less developed, often agricultural, regions in the north and the east to industrial growth centres in the west and south" (Srivastava, 2020a, 165) like Bangalore's expanding tech sector or the Mumbai's finance sector.

In this paper, I locate the struggles of gig workers in the platform economy beyond the platform from the point of production, to the dimension of "platform-adjacents" (Van Doorn & Shapiro, 2023), specifically state-regulated organizing efforts both through formal infrastructures and informal networks created bottom-up. I delineate current debates on the tensions within the platform economy in India, through putting into conversation prominent academic scholarship in 2024, in addition to broader frameworks of analyses developed in response to the pandemic. I utilize the notions of the worker as "informal", "unorganized", and migrant", from current discussions in academia and news media (Carswell et al., 2022; Chauhan, 2024; Katta et al., 2024; Srivastava, 2020; The Wire Staff, 2022; Yadav & Priya, 2021). Following that, I look at the current infrastructure provided the central government for digital workers – the e-Shram Portal. Through the lens of "perceived affordances" (Bucher & Helmond, 2018), and situating worker self-organization in the context of the aftermath of COVID 19 lock-downs, I argue that more work need to be done in publicizing, integrating and expanding similar initiatives this portal at both central and regional levels to combat the pressing challenges posed by platforms today. I conclude by suggesting pathways for future research on the platform economy that can supplement state and smaller-scale collective efforts towards organizing labor and expanding "the range of regulatory and policy responses to platform power" (Van Doorn and Shapiro, 2023, p. 1).

## Research question and methodologies

RQ: What are the *perceived affordances* of the national database portal e-Shram in India?

To chart the current developments in India's platform economy, I start from examining current secondary sources approaching platforms from an anthropological, quite humanistic perspectives, where some argue that platforms are precisely marching on with a completely opposite, techno-utopian, post-anthropocentric ideology towards its resource and human capital. I place in conversation the recent scholarship on the issue, through a critical examination of the analyses and conclusions of research conducted on unorganized laborers, particularly those coming from a migrant background, in urban centers, engaged in low-level service work in the platform economy. I also use recent labor and migration census reports published by the Indian government's central labor and statistics ministry in the aftermath of fallout caused by the COVID lockdowns.

For web-based platforms, two methodological approaches are suitable entry points. Bucher and Helmond conceptualized Norman's theory of "perceived affordances" for digital apps (see Norman, 1990, as cited in Bucher & Helmond, 2018) and Light et. al provide the valuable methodological approach of the "walkhthrough" (2018) for studying the affordances of these apps. Norman's prescriptive idea – that good design should allow users to interact better with the object – laid the general foundation of our current theory of affordances in modern design, later to be adopted by web and app designers. The idea that affordances are perceived first and then operationalized, is a valuable lens for approaching e-governance platforms. What do citizens on these platforms seek – what affordances of that service do they *perceive* that the platform provides? Conceptually and logically this leads to the scrutiny how the affordances actually materialize on these platforms, through the application of more "natively-digital" methods (Rogers, 2024). Light et al. Further point out two ethical considerations for studying apps through the participant-observer method; namely disturbance caused to users through interactions with a research account, and the sensitivity of personal data of users on social media. Precisely, if online data is already public, what are the limitations of re-purposing it for research? Reflecting on this is important not only for working with personal data of random users in social media research, but also in the case of statistical data on e-governance that obfuscates the lines of database-website-platform. I argue that since ministries already release sensitive data through anonymization, it is relatively easier to re-purpose these particular kinds of quantitative data.

For studying web historiography – for example for platforms and portals that are web-based in addition to having a mobile app, Rogers points to the value of the Internet Archive's Wayback Machine, as a tool for creating "screencast documentaries" to study the digital history of web pages and archived media content (2024, p. 49). These methodologies are not fully effective when applied in isolation, rather through supplementing them together, I argue it is possible to gain valuable perspectives on the infrastructures around platforms and gig work.

## The challenges concerning the Indian platform economy today

In India, while discussions on the platform economy have gained significance in recent scholarship, most scholars have studied receptive or representative effects of content platforms, such as social media or over-the-top (OTT) platforms; or on capital-labor relations at the point of production (PoP); the governance of labor forces and regulation of service apps such as food and essential goods delivery, ride-sharing or beauty and healthcare services (Menon, 2024; Pain, 2024; Ray, 2024; Sarkar, 2024; Sinha & Pandit, 2023). Less attention has been paid to the "platform-adjacents" supporting the platform economy (Van Doorn & Shapiro, 2023, p. 3). Scholars have only recently begun to question how the platform economy is structured, and how the labor force is governed on various parameters, over asking simply how labor produces value on these platforms (Katta et al., 2024; Nair, 2023; Thomas, 2024). On the recent transformations of labor conditions in India, both specifically and in the broader context of the Global South, the present scholarship has placed the neoliberal state as the problematic (non)regulator of platform companies in delineating gig work as low-compensated, sub-contracted, and overall informal, while initiating huge efforts in providing informal workers with workshops and training programmes for skill acquisition (Carswell et al., 2022; Nair, 2023; Srivastava, 2020b; Yadav & Priya, 2021).

Organizing workers in India, particularly migrant laborers pursuing upward mobility from the informal sector, has historical posed significant structural and operational challenges. In the face of existential precarity and authoritarian social policies, unorganized gig workers have turned to self-organizing in various major cities (Ray, 2024). While such informal organizing has been instrumental at a certain level in providing migrant workers with security and networking, factors such as caste, and spatial rooted-ness in urban destinations cause barriers to the leverage these informal networks provide certain groups of migrants (p. 1237). The inequalities of the caste structure emerge through forms of labor control in specific regions and urban economies, continuing the exploitative tendencies of capitalism through the platform economy. Non-platform intermediaries and local elites have also been integral to low-level service work; recruitment agencies, skills training programmes supplement migrant workers in their introduction and on-boarding to gig work (Nair, 2023; Ray, 2024; Sinha & Pandit, 2023; Srivastava, 2020; Yadav & Priya, 2021). Ultimately, migrant groups are forced into 'servile docility" (Zhou 2022, as cited in Katta et al., 2024, p. 1106). Migrant labor which was already in the informal sector find themselves disenfranchised by draconian disciplinary mechanisms of platforms, specially through unequal access to social services and deportation (Katta et al., 2024; Srivastava, 2020a, 2020b; Yadav & Priya, 2021).

Due to the (perceived) lack of formal infrastructural systems to support gig workers, the latter are increasingly forming local, informal support networks as a means to survive in the platform economy across urban-rural geography of India, as Ray argues from his interviews with migrant gig workers engaged in ride-hailing and food delivery (2024, p. 1231). The hypocritical behavior of platforms, in projecting their service workers as "heroes" and "guardians" of our society ostensibly backfired when these platforms couldn't provide basic security to the very workers in terms of working conditions, stable wages, occupational security, housing and healthcare (Ray, 2024; Sinha & Pandit, 2023). Indeed, van Doorn and Shapiro's *s*peculation about the behavior of platform companies rang even more true during the turning point of this decade; that "platforms change their story depending on who they are addressing, including financial and market authorities, investors, shareholders, policymakers, courts, and workers." (Van Doorn and Shapiro, 2023, p. 20).

## Case Study: e-Shram and the promise of institutionalization

India presents a unique case for the intricacies between education and employment. According the central government's Periodic Labour Force Survey, most recently from July 2023 – June 2024, indicates that unemployment is estimated at 10.1% - but most surprisingly, it is more (7.9%) for youth aged 15 and above who have secondary schooling or higher levels of education, and less (6.5%) for those without (Ministry of Statistics and Program Implementation, 2024). A second statistic from the recent PLFS report is further concerning: 20.1% of urban women are unemployed vs rural women (8.2%). This breaks every notion eschewed towards the developing Global South from the West that formal education provides pathways towards employment and upward class mobility and points towards the realities of platform capitalism today. Formal education no longer provides safe guarantees of respectable work, and that skills outside the classroom are contributing towards greater number of Indian youth joining the informal economy, which is increasingly permeating into the platform economy.

On 26th 2021, the central government launched the *e-Shram* Portal, which is stated as a comprehensive National Database of Unorganized Workers (NDUW) launched by the central government of India, under the labor ministry. The front page of the portal (Figure 1 and 2) points towards specific affordances – A sub-portal for jobs, skills, and tabular data of the registered unorganized workers, along with a direct link towards pension registration on *maandhan.in*. A press briefing, dated 3 years later, summarizes detailed promises of the scheme: universal account numbers (UAN), e-Shram identity cards, and easier access to existing and proposed government schemes, services and facilities (Himanshu Pathak, 2024). The briefing states that as of 31st July, 2024, 29.85 crore (over 298 million) unorganized workers had registered (2024).

**Figure 1**

*Home page of the e-Shram portal*

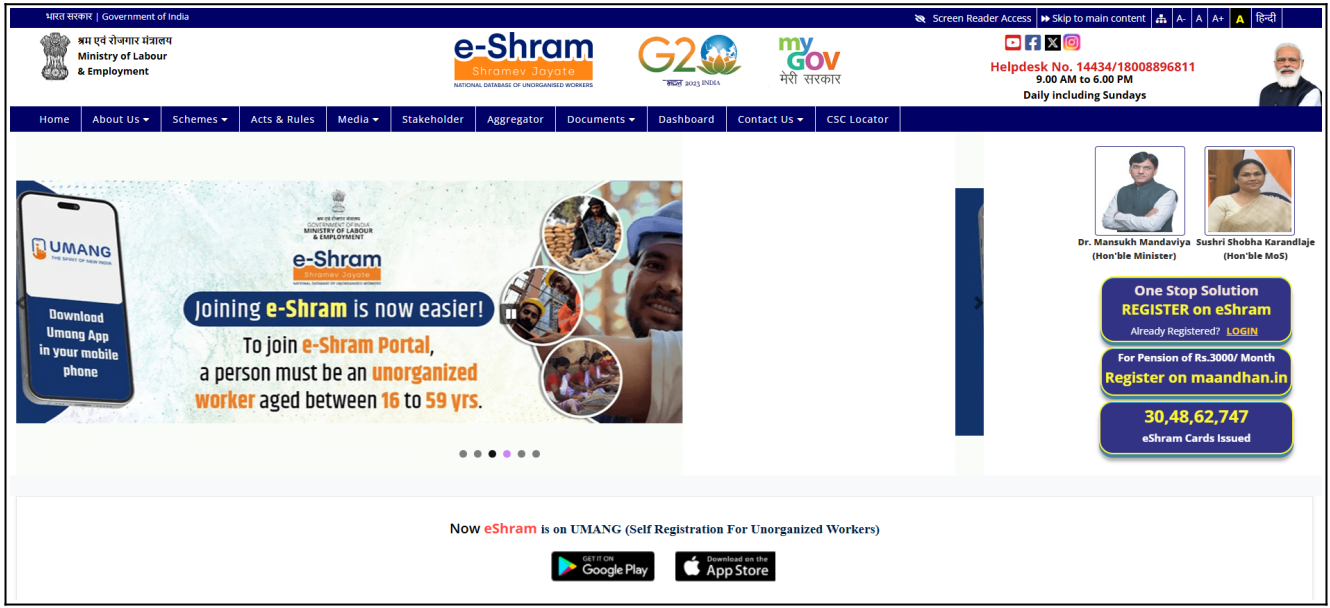

**Figure 2**

*Affordances on the main page of e-Shram*

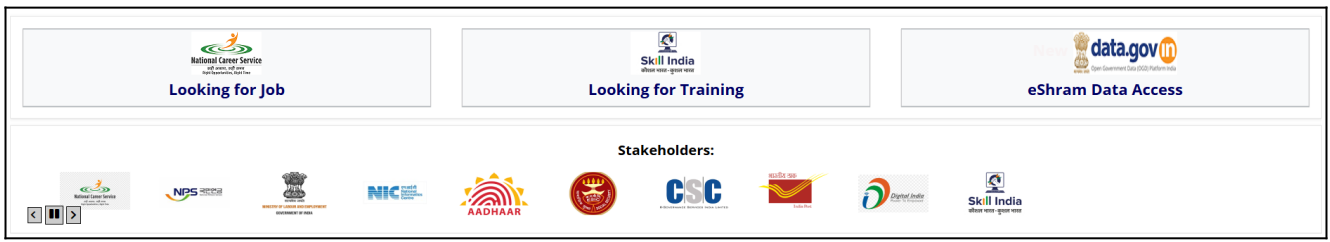

**Figure 3**

*Appification*

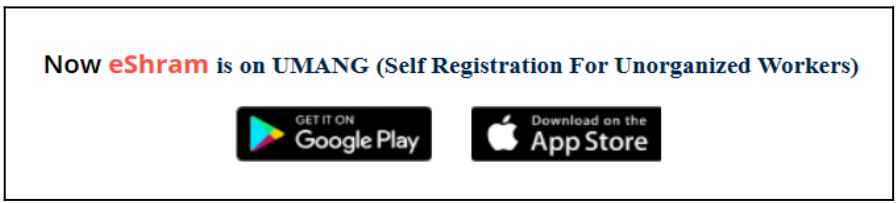

A banner also says that the portal is available on the mobile app UMANG (Figure 3), available on the standard app stores. Appification and the reliance on the infrastructures of Google Play and Apple's App Store complicates the structural concerns with the platform, particularly in relation to security of the database against extractive practices of monopolistic tech companies contending with national regulations for maximization of their shareholder profits.

**Figure 4**

*Redundancy in occupational categories*

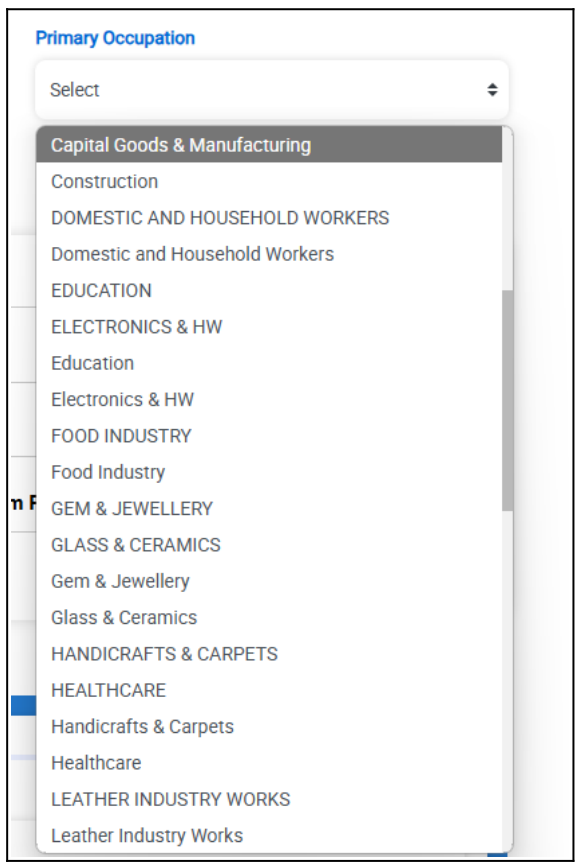

Furthermore, I found a very critical, and debilitating error in the database upon further inspection through the Data Access Page, which gives individual anonymized records of unorganized workers in each sector. When I was exploring the filters and categories in the input sections to get the tabular data, I found that in the 'Primary occupation' category, a lot of crucial sectors have redundant instances, for example with different capitalization in syntax. The repeating categories discovered while searching sectorial data for West Bengal were –

'Wood & Carpentry', 'Leather Industry Works', 'Food Industry', 'Domestic and Household Workers', 'Electronics & HW'. This redundancy already obfuscates researchers and policy makers trying to access the data, notwithstanding problems arising from how the recording, assemblage and classification of the data occurred at the server-end for some of the most crucial informal sectors like food and domestic work.

Running the Wayback Machine, I found that the portal was only archived twice before, 13[th] and 16[th] December, 2024 (the 19[th] December access log is the methodological instance of this paper), while site has been operational since 26[th] August 2021. This raises the immediate question of why this platform did not received particular attention during its developmental stages, more so during the early days of its deployment. A few news pieces were published by digital publications like *Scroll.in* discussing *how* unorganized workers received this new infrastructure (Chauhan, 2024), and how they perceived this would help in alleviating their daily struggles under platform power. Generally, reception among unorganized workers was starkly negative compared to how the government praised its effectiveness in the Lok Sabha.

**Figure 5**

*Snapshot history in Wayback Machine*

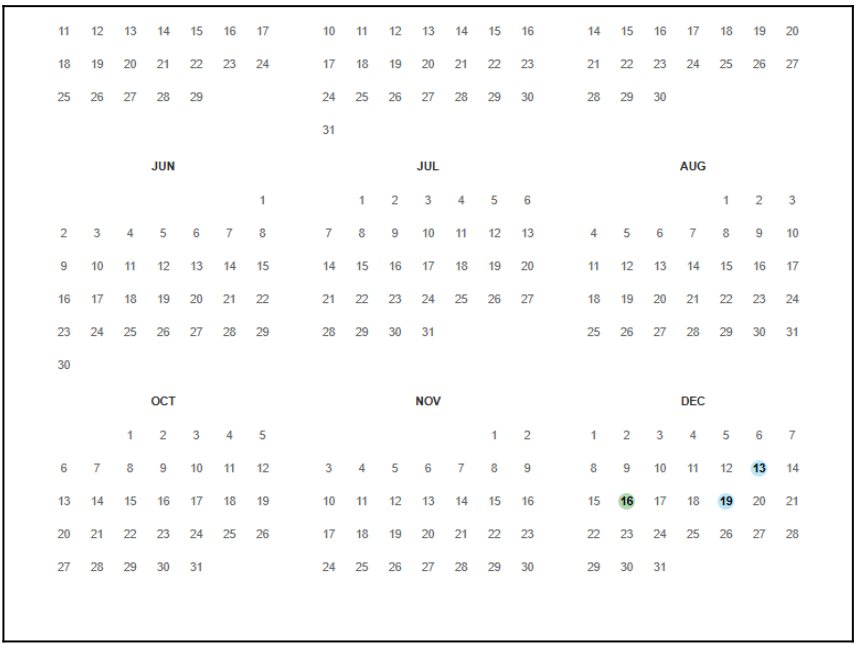

The 298 million registrations publicized widely by the central government in parliamentary proceedings and through corporate media and business manifests fails to underscore the stark contrast in ground realities: India's informal labor sector consisted around 439 million workers in 2019-20, according to the Economic Survey of 2021-22 (Chauhan, 2024). In 3 years, the efforts of the central government has realistically registered around half of the informal sector, bearing mind the influx of informal workers into the restructured platform economy following COVID lockdowns. This under-representation of unorganized workers has been attributed to issues with scaling up registration, digital inaccessibility, and other technological barriers (Chauhan, 2024; Srivastava, 2020).

## Discussions and conclusion

In this paper, I have attempted to constructively deviate the attention of scholarship on the Indian platform economy away from issues of representation and globalization on and through content platforms, and also broadly discussed issues of platform power and algorithmic control that have risen in the case of service-oriented platform work. Reflecting on current scholarship on the Indian platform economy, I identified certain gaps, namely the lack of research on e-governance platforms that support unorganized migrant workers, who are often employed informally even in platform regimes. I explored the database-portal-platform *e-Shram*, which was deployed post-COVID lock-downs in 2020 by the central government to provide unorganized migrant workers with a platform for registration, identification, and easier access to government social security schemes both at state and local levels. I argue that initiatives like this firstly need to integrated better, secondly marketed more broadly owing to the linguistic diversity of the country, and finally be operationalized as a compliment to platform networks and informal support networks of socio-spatial labor groups. The regulators, the state and the platform companies, must be pushed to address the gaps between what platforms *do* and what they *say* they do (Van Doorn and Shapiro, 2023, p. 20) – so that gig workers find non-restrictive pathways towards security in their work. Capital, here information capital specifically, has already helped in deploying these platforms to an extent through extractive practices and the data-center industrial complex. Now labor, and platform-adjacents like state authority and even non-platform intermediaries must renegotiate their presence within these platforms as infrastructural, rather than solely through its relation with the data capital of these platforms. Prosthetic solutions like e-Shram are beneficial, but need to be scaled up and fixed to reflect meaningful impact.

I am personally interested in continuing research in this direction, through broader ethnographic methodologies, although I understand the limitations, and certain sensitivities of conducting telephonic interviews overseas in the Global South while working in Western academia. Furthermore, NDAs might often restrict sub-contracted platform workers from going into intricacies of their work situations and daily grapples with platform power. A more suited approach is supplementing ethnographic methods with being a 'participant-observer' in an example platform, which could be possible during long-term field work in India. A third approach could be studying the self-representation of migrant workers on social media, as some scholarship in other national platform economies has recently shown (Orth, 2024; Pires et al., 2024; Qadri & D'Ignazio, 2022; Qadri, 2021; Rodríguez-Modroño et al., 2024), by launching from the background of COVID restrictions and the way it restructured India's informal sector into platformized services, and taking into account more specific regional contexts currently influencing economic and political decisions in the bureaucracy, like recent floods in Assam and Bengal, a massive landslide in Kerala and ethnic conflicts in Manipur that are threatening the destabilization in the eastern-most region of the country, a region which has historically been a large supply of migrant labor in the circular labor economy.

# Note

The author would like to thank Sushovan Dhar, Rudranil Raha, and Dr. Niels van Doorn for their valuable insights and discussions on the research issue.